\documentclass[]{article}
\usepackage{spconf}

\usepackage{amsmath,amsfonts,amssymb}
\usepackage{graphicx}
\usepackage{subcaption}
\usepackage{adjustbox}
\usepackage[colorlinks=true, allcolors=blue]{hyperref}
\usepackage{xcolor}
\usepackage{wrapfig}

\title{Anisotropic Super Resolution in Prostate MRI using Super Resolution Generative Adversarial Networks}
\name{Rewa Sood \textsuperscript{1}, Mirabela Rusu \textsuperscript{2}}
\address{1. Electrical Engineering, Stanford University  2. Radiology, Stanford University}

\begin{document}
\maketitle

\begin{abstract}
Acquiring High Resolution (HR) Magnetic Resonance (MR) images requires the patient to remain still for long periods of time, which causes patient discomfort and increases the probability of motion induced image artifacts. A possible solution is to acquire low resolution (LR) images and to process them with the Super Resolution Generative Adversarial Network (SRGAN) to create a super-resolved version. This work applies SRGAN to MR images of the prostate and performs three experiments. The first experiment explores improving the in-plane MR image resolution by factors of 4 and 8, and shows that, while the PSNR and SSIM (Structural SIMilarity) metrics are lower than the isotropic bicubic interpolation baseline, the SRGAN is able to create images that have high edge fidelity. The second experiment explores anisotropic super-resolution via synthetic images, in that the input images to the network are anisotropically downsampled versions of HR images. This experiment demonstrates the ability of the modified SRGAN to perform anisotropic super-resolution, with quantitative image metrics that are comparable to those of the anisotropic bicubic interpolation baseline. Finally, the third experiment applies a modified version of the SRGAN to super-resolve anisotropic images obtained from the through-plane slices of the volumetric MR data. The output super-resolved images contain a significant amount of high frequency information that make them visually close to their HR counterparts. Overall, the promising results from each experiment show that super-resolution for MR images is a successful technique and that producing isotropic MR image volumes from anisotropic slices is an achievable goal.  
\end{abstract}


\section{INTRODUCTION}
\begin{figure*}[h!]
    \centering
    \includegraphics[width=4.5in]{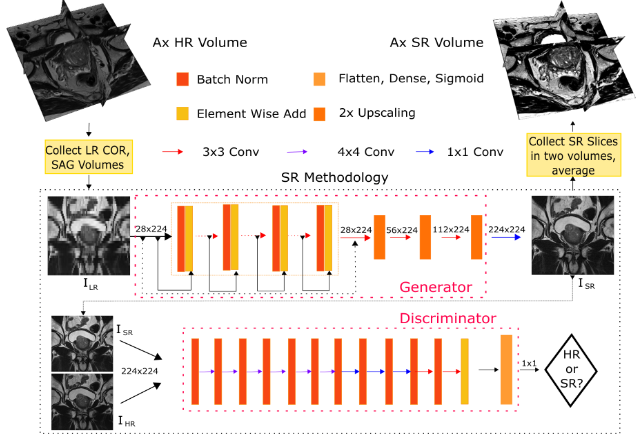}
    \caption[First figure]{SRGAN architecture: The SRGAN is composed of a generator and discriminator network trained in an adversarial fashion and processes 2D images but creates a 3D volume by stacking SR outputs}
    \label{fig:SRGANarch}
\end{figure*}

Acquiring high-resolution (HR), clinically usable MR images is time consuming, expensive, and uncomfortable for the patient. An increase in scanner throughput can be achieved by acquiring low-resolution (LR) images instead of HR images and subsequently post-processing them to form super-resolved (SR) images of the same perceptual quality as the original. Recently, Ledig et. al. proposed the Super Resolution Generative Adversarial Network (SRGAN) which uses a perceptual loss that produces the visually pleasing results in natural images ~\cite{DBLP:journals/corr/LedigTHCATTWS16}. The SRGAN architecture consists of two parts: the generator and discriminator (Figure \ref{fig:SRGANarch}). The downsampled versions of the HR in-plane slices are extracted from the MR volume and fed as batches into the network. The generator is trained to fool the discriminator into believing that the output SR images are HR, while the discriminator is trained to distinguish SR images from their HR counterparts. The GAN approach uses a loss function that is comprised of a perceptual loss, which encourages SR reconstructions to move towards regions of the search space with high probability of containing photo-realistic images and a content loss based on perceptual similarity using the high-level features from a pretrained VGG19 network~\cite{DBLP:journals/corr/JohnsonAL16}. This work proposes the use of the SRGAN to produce SR versions of LR MR images. The original SRGAN implementation is modified in three ways. The SRGAN network is first changed to work with grayscale MR prostate images to produce both a 4 and $8\times$ isotropic increase in in-plane resolution. Second, anisotropic super-resolution is performed using images that are synthetically created by downsampling only the HR image height. Finally, the network is applied to through-plane slices, which have a natural anisotropic resolution due to the lower through-plane resolution inherent in MR images.

\section{RELATED WORK}
Traditionally,  SR in medical imaging has been achieved through image processing techniques or modifications to the image acquisition protocol. For example, the algorithm in Rousseau et.al.~\cite{priors} uses anatomical intermodality priors from a reference image, while in Peeters et. al.~\cite{fmri}, the authors collected slice-shifted images which they interpolated to decrease the effective slice thickness and increase the SNR. Recently, many authors have applied machine learning based techniques to the medical image SR problem. Both Yang et.al.~\cite{representation} and Park et.al.~\cite{deepcnn} use convolutional neural networks to achieve SR. While the previous works attack the SR problem from the 2D perspective, others such as Chaudhari et.al.~\cite{doi:10.1002/mrm.27178} and Chen et.al.~\cite{3dbrain}, form 3D solutions. As analyzed in the authors' previous paper ~\cite{prev}, the machine learning based techniques outperformed the traditional techniques in both the quality of the SR images and in the amount of time required to acquire the result. However, it was also shown that a machine learning implementation using GANs produced the best visual results over all. Li et.al.~\cite{thinslice} use GANs with 3D convolutional kernels to reduce the slice thickness of the aquired MR volume. While this approach produced better results than 2D and 3D SRCNN, training a 3D network has limitations that a 2D network does not face. Training a 3D network requires more compute because of the significant increase in parameters. Additionally, the amount of data available decreases because the input volume as a whole represents one data point instead of each slice. This work circumvents using a 3D network by approaching SR per slice and reconstructing the MR volume at the end, while still providing results that are visually close to the ground truth. 
    
\section{METHODS}

We use the Prostate-Diagnosis~\cite{CIAWikiDiagnosis} and PROSTATEx~\cite{CIAWikiChallenges} datasets from the Cancer Imaging Archive. Prostate-Diagnosis contains acquisitions for 87 patients while the PROSTATEx dataset contains data for 242 patients. Data from 82 patients in Prostate-Diagnosis and 238 patients in the PROSTATEx dataset were included in the training set while the remaining 9 were included in the test set. A validation set was not used in this work because of the small amount of data involved. Instead, a batch of training images per epoch was used for online validation. The DICOM images from these datasets are first converted to PNG and scaled. We run all of the ML SR techniques using an NVIDIA Tesla K80 GPU.  

\begin{figure*}[h!]
    \centering
    \begin{subfigure}[h!]{0.26\textwidth}
        \includegraphics[height=0.9in,trim=30 100 30 75,clip]{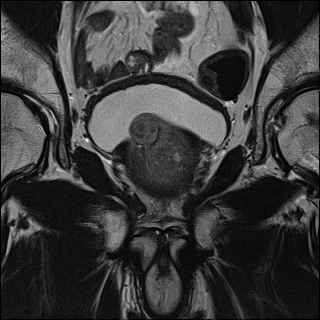}
        \caption{HR Ground Truth}
    \end{subfigure}%
    \begin{subfigure}[h!]{0.24\textwidth}
        \includegraphics[height=0.9in,trim=25 70 25 55,clip]{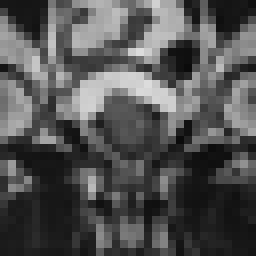}
        \caption{8x LR input}
        \label{LRINP}
    \end{subfigure}%
    \begin{subfigure}[!h]{0.25\textwidth}
        \includegraphics[height=0.9in,trim=30 100 30 75,clip]{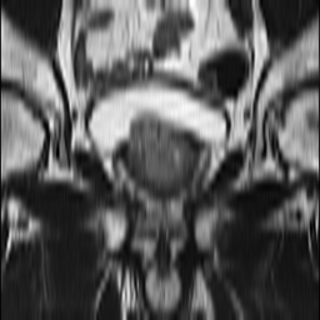}
        \caption{Bicubic Interpolation}
    \end{subfigure}
    \begin{subfigure}[h!]{0.24\textwidth}
        \includegraphics[height=0.9in,trim=30 100 30 75,clip]{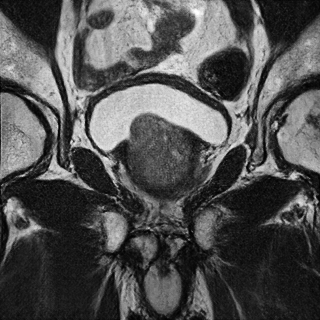}
        \caption{SRGAN 8x Output}
    \end{subfigure} 
    \caption{LR input to and SR output of synthetic isotropic SRGAN model for 8x upscaling}
    \label{fig:SRGANComparison}
\end{figure*}

\begin{figure*}[h!]
    \centering
    \begin{subfigure}[h!]{0.26\textwidth}
        \centering
        \includegraphics[height=0.9in,trim=25 80 25 60,clip]{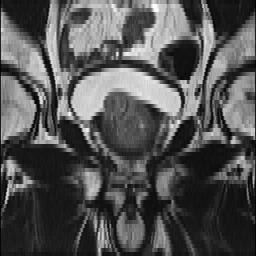}
        \caption{Input LR, 4x}
    \end{subfigure}%
    \begin{subfigure}[h!]{0.24\textwidth}
        \centering
        \includegraphics[height=0.9in,trim=25 80 25 60,clip]{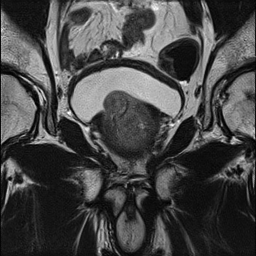}
        \caption{Output SR, 4x}
    \end{subfigure}%
    \begin{subfigure}[h!]{0.25\textwidth}
        \centering
        \includegraphics[height=0.9in,trim=20 70 20 55,clip]{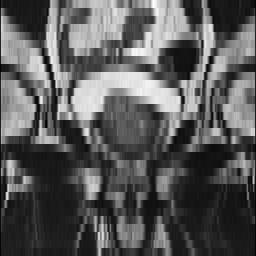}
        \caption{Input LR, 8x}
    \end{subfigure} 
    \begin{subfigure}[h!]{0.24\textwidth}
        \centering
        \includegraphics[height=0.9in,trim=25 80 25 60,clip]{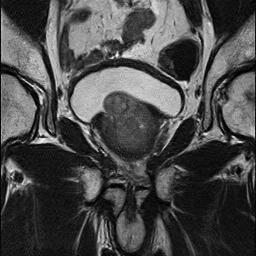}
        \caption{Output SR, 8x}
    \end{subfigure}
    \caption{LR input to and SR output of synthetic anisotropic SRGAN model }
    \label{fig:AnisoComparison}
\end{figure*}

The SRGAN network is executed using an Adam optimizer with a batch size of 16, learning rate of 0.0001 and beta1 of 0.9. The generator is executed standalone for 20 epochs and then the discriminator and generator are trained sequentially for 50 epochs. The generator has the difficult task of creating new edge information, while the discriminator only has to solve the simple classification problem of whether a particular image is HR or SR. Thus, training the generator network on its own for 20 epochs ensures that the discriminator will not dominate the learning process and preclude the generator from learning anything at all. Overall, the inputs to this network are 28x28 or 56x56 LR images and the outputs are the 224x224 SR versions.

Using the same training protocol and hyperparameters as Experiment 1, Experiment 2 tests the ability of the SRGAN to provide SR results for anisotropic resolutions, where the physical pixel size differs between the height and width of an image. The block in the original network that produces the isotropic upscaling is modified to upscale only the height dimension of the input image. The input images are cropped to 224x224, but are downsampled to 28x224 (height x width). The output images are the 224x224 SR versions.

This final experiment builds on the previous one by using real anisotropic images instead of the synthetic version. For a 320x320x26 (height x width x depth) volume, the 320 320x26 (height x depth) and (width x depth) through-plane slices are collected and saved as PNGs. For example, in an axial volume these through-plane slices are the LR sagittal and coronal perspectives. The two distinct sets of through-plane images are used as separate inputs to the anisotropic SRGAN. The output volumes are constructed by stacking the output images. The final SR volume is taken as the average of the volumes produced by each through-plane set. 

\section{RESULTS}

\begin{table*}[!h]
	\centering
    \begin{adjustbox}{width=0.90 \textwidth,center=\textwidth}
	\begin{tabular}{| l | l | l | l | l | l |  l |}
    \hline
    & 4x Bicubic (Iso) & SRGAN4x & SRGAN8x & 8x Bicubic (Aniso) & Anisotropic 4x & Anisotropic 8x  \\
    \hline
    PSNR [dB] & 21.68 & 21.27 & 18.73 & 25.72 & 29.51 & 25.72 \\
    SSIM & 0.71 & 0.66 & 0.47 & 0.76 & 0.82 & 0.70 \\
    \hline
    \end{tabular}
    \end{adjustbox}
    \caption{SR performance results: The PSNR and SSIM metrics do not reflect the perceptual quality of the images as they are higher for bicubic interpolation than the GAN methods.}
    \label{table:performance_results}
\end{table*}

Figure \ref{fig:SRGANComparison} contains an example SR output for the SRGAN 8x network and the associated LR, bicubic interpolation, and HR images. The LR input image (Figure \ref{LRINP}) is severely pixelated and has no edge fidelity. While the image produced via bicubic interpolation has no pixelation, this method is still unable to preserve the high frequency information found in the ground truth image. In the SRGAN 4x and 8x models, the discriminator network seeks out the high frequency information that differentiates HR and LR images, thus forcing the SRGAN output to have high frequency details. The SRGAN 8x network is not able to maintain as high an edge fidelity as the SRGAN 4x network. This result is expected because the SRGAN 8x network is provided with far less information since the input LR image is a further 2x smaller in both dimensions. In the LR images in Figure \ref{fig:AnisoComparison}, the horizontal dimension has the same resolution as that of the HR image while the resolution of the vertical dimension is a factor of 4 or 8 less. The modification to the upscaling block in the SRGAN model allows the network to super-resolve images in an anisotropic fashion. The output SR images are not pixellated or blurred and are visually close to the HR ground truth image.

\begin{figure*}[h!]
    \centering
    \begin{subfigure}[h!]{0.33\textwidth}
        \centering
        \includegraphics[height=0.85in,trim=0 80 0 70,clip]{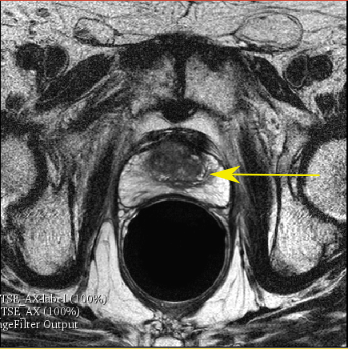}
        \caption{HR axial slice}
    \end{subfigure}%
    \begin{subfigure}[h!]{0.33\textwidth}
        \centering
        \includegraphics[height=0.85in,trim=0 30 0 10,clip]{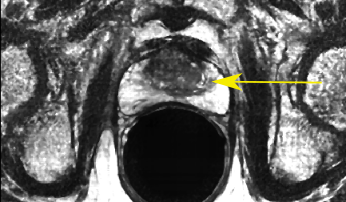}
        \caption{SR axial slice}
    \end{subfigure}%
    \begin{subfigure}[h!]{0.33\textwidth}
        \centering
        \includegraphics[height=0.85in]{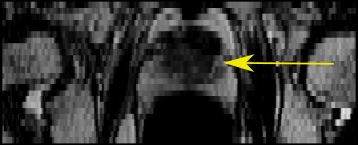}
        \caption{LR axial slice}
    \end{subfigure}%
    
    \begin{subfigure}[h!]{0.32\textwidth}
        \centering
        \includegraphics[height=0.85in,trim=5 20 5 20,clip]{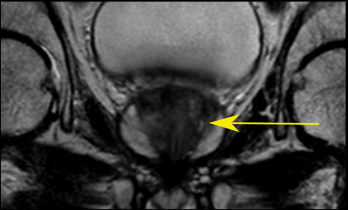}
        \caption{HR coronal slice}
    \end{subfigure} 
    \begin{subfigure}[h!]{0.32\textwidth}
        \centering
        \includegraphics[height=0.85in,trim=0 20 0 20,clip]{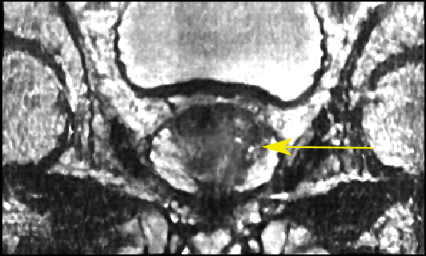}
        \caption{SR coronal slice}
    \end{subfigure}
    \begin{subfigure}[h!]{0.32\textwidth}
        \centering
        \includegraphics[height=0.85in,trim=0 20 0 20,clip]{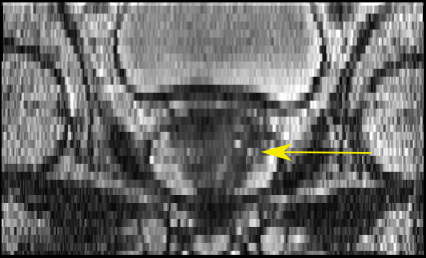}
        \caption{LR coronal slice}
    \end{subfigure}
    
    \caption{SR result for anisotropic SRGAN using real data, yellow arrow indicates suspicious cancer region}
    \label{fig:VolumeAnisoComparison}
\end{figure*}

Table ~\ref{table:performance_results} describes the PSNR and SSIM results for the 4 and 8x isotropic and anisotropic SRGAN implementations compared to their respective bicubic interpolation baselines. PSNR measures the signal to noise ratio in the image, while SSIM measures the correlation between two images and ranges from -1 to 1, with -1 for opposite images and 1 for identical images. The PSNR and SSIM results for the 4x and 8x confirms the expected qualitative result above that the SRGAN 8x network is unable to create the same amount of high frequency information as the 4x network. Additionally, both of these networks have worse quantitative results than the bicubic interpolation results for two main reasons. First, bicubic interpolation provides a smoother SR result than either isotropic SRGAN method because it minimizes the MSE, which leads to a higher PSNR. Additionally, the SRGAN works to reproduce edge information, and while it succeeds in doing so, the edges produced may not be exactly the same as the edges in the HR image, which leads to a lower SSIM result. However, both the isotropic and anisotropic methods have at best similar quantitative results to their bicubic interpolation baselines, indicating that the PSNR and SSIM metrics are not sufficient to quantify the demonstrated improvements in edge fidelity.  

Figure \ref{fig:VolumeAnisoComparison} is an example of axial and coronal views. The HR views portray the difference in resolution between the in-plane axial slice and the coronal and sagittal through-plane slices. The SR views show the anisotropic SRGAN outputs on the real HR data. The arrows in the top row of Figure \ref{fig:VolumeAnisoComparison} point to a suspicious region within the axial prostate. The LR axial image obtained by taking a through-plane slice from a coronal volume has lost all information about this region. The SR image is able to reconstruct the high frequency content and provides valuable edge and texture information. The anisotropic SRGAN method is able to improve the through-plane resolution by 8x while incurring only a slight resolution loss in the in-plane slice.   

\section{CONCLUSION}
Each experiment has been successful in showing that the SRGAN can be used to super-resolve prostate MR images. The first experiment demonstrated that the network was able to reproduce high frequency information via isotropic SR. The second experiment showed that the SRGAN is not constrained to isotropic SR, while the third experiment proved that the SRGAN can produce 8x improvements in the through-plane resolution of the prostate MRI. This method could be useful to physicians by allowing a shorter scan time while still providing a clear representation of the prostate, such as in the apex which has a higher probability of containing cancer \cite{rusu}. Additionally, these promising results are significant steps toward creating a fully isotropic prostate MRI, which can improve the results of registration and segmentation tasks by providing more accurate edge information.

\bibliography{report}
\bibliographystyle{spiebib}

\end{document}